\documentstyle[prl, aps, epsf, multicol]{revtex}
\newcommand{\NP}{Nucl. Phys. }
\newcommand{\PR}{Phys. Rev. }
\newcommand{\PRL}{Phys. Rev. Lett. }
\newcommand{\PL}{Phys. Lett. }

\begin{document}

\title{\bf Non-unitarity of CKM matrix from vector singlet quark mixing
and neutron electric dipole moment
}
\author{Yi Liao\footnote{Present mailing address:
Institut f\"ur Theoretische Physik, Universit\"at Leipzig,
Augustusplatz 10/11, D-04109 Leipzig, Germany}}
\address{Department of Physics, Tsinghua University,
Beijing 100084, P.R.China}
\author{Xiaoyuan Li}
\address{Institute of Theoretical Physics, The Chinese Academy of Sciences,
Beijing 100080, P.R.China}

\maketitle

{\tighten
\begin{abstract}

In the standard model (SM) the lowest order contribution to the quark
electric dipole moment (EDM) occurs at the three loop level. We show
that the non-unitarity of the CKM matrix in models with an extended
quark sector typically gives rise to a quark EDM at the two loop level
which has no GIM-like suppression factors except the external quark
mass. The induced neutron EDM is of order $10^{-29}$ ${\rm e~cm}$ and
can be well within the reach of the next generation of experiments
if it is further enhanced by long distance physics as happens in the SM.

\vspace{0.2cm}


\end{abstract}
}
\vspace{0.3in}
\begin{multicols}{2}\narrowtext


An important target of particle physics is the determination of the
Cabibbo-Kobayashi-Maskawa (CKM) matrix\cite{ckm}, which parameterizes
the charged current interactions of quarks. In the standard model (SM)
with three generations of quarks the CKM matrix is a $ 3 \times 3 $
unitary matrix, and the $CP$ violation is due to the presence of a
nonzero phase in this matrix. The unitarity of the CKM matrix is
essential in suppressing flavour changing neutral current (FCNC)
processes\cite{GIM}\cite{GW}. Using the CKM unitarity relation the $CP$
violating phase information can be elegantly displayed in terms of
unitary triangles. The most interesting challenge of the $B$-factories
now entering into operation, and of future collider $B$ experiments
is to try to pin down the angles in the unitary triangles\cite{peccei}.
If the measured angles violate either of the `triangle conditions', or
correspond to a point $(\rho, \eta) $ \cite{wolf} which is outside the
allowed region, then we will have evidence for new physics.

Violation of the CKM unitarity can appear in models with an extended
quark sector\cite{nir}. Various constraints on the possibility that
exotic quarks mix with the ordinary SM quarks have been derived
from low energy charged and neutral current phenomenology, $Z$ physics,
FCNC processes and $CP$ violation in neutral
meson systems\cite{zphys}\cite{bbbar}\cite{bsgamma}. In this letter
we propose to examine the unitarity of the CKM matrix in a different
setting, namely, by investigating possible unitarity violating effects
on the neutron electric dipole moment (EDM). We shall find that
information from the neutron EDM is complementary to that from FCNC
processes in $B$ physics and serves as a self-consistency check of the
relevant theory.

In the minimal SM quark EDM's vanish at the one loop level because the
relevant amplitudes do not change the quark flavor and each CKM matrix
element is accompanied by its complex conjugate so that no $T$-violating
complex phase can arise. At the two loop level individual diagrams can
have a complex phase, but it has been shown by Shabalin\cite{shab} that
their sum vanishes strictly. The null result was confirmed afterwards
by several groups\cite{khrip}. It is thus thought that in the
SM the lowest order contribution to quark EDM's occurs at the three
loop level. A recent calculation\cite{ck97} shows they are of order
$10^{-35}$ to $10^{-34}$ e cm for $u$ and $d$ quarks.
The extreme smallness of the quark and neutron EDM's in the SM makes them
particularly suited for searching for new physics. The current
experimental upper bound\cite{neutron}\cite{recent},
$|d(n)|< 6.3\times 10^{-26}{\rm e~cm}$,
has put very strigent constraints on extensions of the SM,
such as additional Higgs fields, right-handed currents, or
supersymmetric partners\cite{review}. We reanalyzed the problem in
Refs.\cite{ll99} and \cite{ll00} and found that the complete two loop
cancellation can be attributed to two special features in the SM:
the purely left-handed structure of the charged current and the
unitarity of the $3\times3$ CKM matrix. The cancellation introduced by
the unitarity is the flavour diagonal analog of the GIM suppression
in the FCNC processes, with the mere difference being that the
cancellation is complete in the EDM case. Thus it is expected that in
models with an extended quark sector the contributions to quark EDM's
at the two loop level are no longer cancelled thoroughly because of
violation of the CKM unitarity, and that potentially large EDM's for
quarks and the neutron can then be induced. Ignoring possible
logarithmic factors, they are of order,
$\displaystyle d(q) \sim eg^4\pi^{-4}\tilde\delta m_q/m_W^2$,
where $g$ is the semi-weak coupling, $\tilde{\delta}$ is the rephasing
invariant measure of $CP$ violation\cite{jarlskog}. Note that there
are no GIM-like suppression factors except the external light quark
mass which is required by chirality flip.
Numerically they are of order $ 10^{-29}$ ${\rm e~cm}$, well within
the reach of the next generation of experiments\cite{future}
if they are further enhanced by long distance physics as happens in
the SM.

We consider here a model with one extra singlet down-type quark in a
vector-like representation of the SM gauge group,
$SU(3)_C \times SU(2)_L \times U(1)_Y $. In addition to the three quark
generations, each consisting of the three representations$(i=1,2,3)$
\begin{equation}
Q^i_L(3,2)_{+1/6},~~u^i_R(3,1)_{+2/3},~~d^i_R(3,1)_{-1/3},
\end{equation}
we have the following vector-like representation:
\begin{equation}
d_4(3,1)_{-1/3}+\bar{d}_4(\bar{3},1)_{+1/3}
\end{equation}
Such a quark representation appears, for example, in $E_6$ GUTs
\cite{rosner}. The model can be considered as a minimal extension of
the SM in the sense that there is no other change in the gauge and
scalar sectors. In particular, the charged currents remain purely
left-handed and the $CP$ violation is still encoded in the CKM matrix.

After spontaneous symmetry breaking, the down-type singlet quark
($d_4$) mixes with the ordinary three down-type quarks so that the
weak and mass eigenstates are related by a $4\times4$ unitary matrix,
\begin{equation}
\left(
\begin{array}{c}d^{\prime}\\s^{\prime}\\b^{\prime}\\d^{\prime}_4
\end{array}\right)_L=
\left(\begin{array}{cccc}
V_{ud}&V_{us}&V_{ub}&V_{u4}\\
V_{cd}&V_{cs}&V_{cb}&V_{c4}\\
V_{td}&V_{ts}&V_{tb}&V_{t4}\\
V_{0d}&V_{0s}&V_{0b}&V_{04}
\end{array}\right)
\left(\begin{array}{c}d\\s\\b\\d_4\end{array}\right)_L.
\end{equation}
In the basis where the up-type quarks are diagonalized, the submatrix
consisting of the first three rows in the above matrix appears in
$SU(2)_L$ $W$ and $Z$ couplings, and it is the generalized CKM matrix
in this model. Note that there still exist unitarity relations among
the up-type quarks although the matrix is no longer unitary. The
non-unitarity of the matrix leads to FCNC $Z$ couplings amongst
down-type quarks which have important repercussions on FCNC and $CP$
violating processes\cite{bbbar}\cite{bsgamma}.
In this work we shall focus on the purely charged current contributions
to the EDM's. We shall also discuss briefly the more complicated
contributions involving FCNC interactions.

\begin{figure}[!b]
\vspace*{-3.0cm}
  \centerline{\epsfxsize=14.5truecm \epsfbox{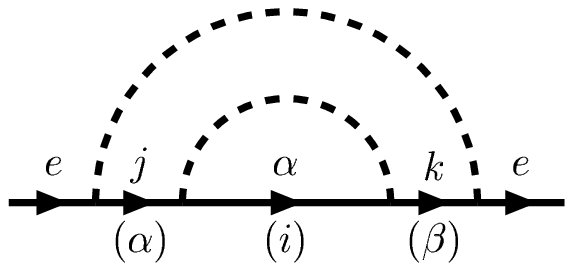}  }
 {\tighten
\caption[1]{ The Feynman diagram that contributes to the EDM of the
up-type quark $u_e$. A background electromagnetic field is understood
to be attached to internal lines in all possible ways. The dashed
lines represent $W^{\pm}$ and $G^{\pm}$ bosons. The diagram for the
down-type quark $d_e$ is obtained by substitutions:
$\alpha\to i$ and $j,~k\to\alpha,~\beta$.}
\label{C0b} }
\end{figure}

Let us consider the quark EDM\cite{singletEDM}. The effective
Lagrangian for the EDM interaction is defined as
$\displaystyle {\cal L}_{\rm eff}=
-id/2\bar{\psi}\gamma_5\sigma_{\mu\nu}\psi F^{\mu\nu}$,
where $F^{\mu\nu}$ is the electromagnetic tensor and $d$ is the EDM
of the fermion $\psi$. Again, no $T$-violating complex phase can arise
at the one loop level. The contributing Feynman diagram at the two
loop level is shown in Fig.~\ref{C0b}.
Following our previous works\cite{ll99}\cite{ll00}, we use the
background field\cite{abbott}-\cite{ll95}
( or the nonlinear\cite{nonlinear} )
$R_{\xi}$ gauge with $\xi=1$. In this gauge, there is no
$W^{\pm}G^{\mp}A$ coupling and the $W^+W^-A$ coupling is also very
simple\cite{ll99}.[ $A$ is the background electromagnetic field and
$G^{\mp}$ are would-be Goldstone fields. ] We use Greek and Latin
letters to denote up- and down-type quarks respectively, and the
external quark is denoted as $u_e$ or $d_e$. There are four groups
of contributions, denoted as $WW$, $WG$, $GW$ and $GG$, where the
first and second letters refer to the bosons exchanged in the outer
and inner loops respectively.

Consider for example the $u_e$ quark EDM. We found as in the SM
\cite{ll99} that the contribution from $WW$ is automatically cancelled
without using any unitarity conditions while other contributions have
the following separate structure:
\begin{equation}
\displaystyle V_{ek}V^*_{\alpha k}V_{\alpha j}V^*_{ej}
[H_{\alpha}(k)-H_{\alpha}(j)],
\end{equation}
where $H_{\alpha}(k)$ is a function of masses of $u_{\alpha}$, $d_k$,
$u_e$ and $W^{\pm}$. The crucial point is that the dependence on $d_j$
and $d_k$ masses splits up. This is mainly responsible for the
thorough or partial cancellation occurring in the SM and beyond.
Summing up the pair of mirror-reflected diagrams
(i.e., $j\leftrightarrow k$) doubles the imaginary part of the product
of CKM matrix elements while removing its real part. The summation
over six pairs of $(jk)$, since we have four down-type quarks
$d_i$, $d_j$, $d_k$ and $d_l$,
completely cancels contributions amongst themselves due to the
unitarity of the $4\times4$ matrix $V$. For example, the
$H_{\alpha}(j)$ term is
\begin{equation}
\begin{array}{rl}
&2i{\rm Im}[V_{ej}V^*_{\alpha j}(V^*_{ek}V_{\alpha k}+V^*_{el}V_{\alpha l}
+V^*_{ei}V_{\alpha i})]\\
=&2i{\rm Im}[V_{ej}V^*_{\alpha j}(\delta_{e\alpha}-
V^*_{ej}V_{\alpha j})]=0.
\end{array}
\end{equation}
In other words, the up-type quark EDM's vanish strictly in the model
with an extra down-type singlet quark as in the SM.  However, for the
down-type quark EDM's in the model, we have one less up-type quark
in virtual loops to complete the unitarity cancellation so that the
cancellation is not thorough. For example, summing over the pairs of
$(\alpha\beta)$ in the contribution to the $d_e$ quark EDM (see Fig. 1
for notations) and using the unitarity of $V$, the $H_i(\alpha)$ term
is
\begin{equation}
\begin{array}{rl}
&2i{\rm Im}[V^*_{\alpha e}V_{\alpha i}(V^*_{\beta i}V_{\beta e}+
V^*_{\gamma i}V_{\gamma e})]\\
=&2i{\rm Im}[V_{\alpha e}V^*_{\alpha i}V_{0i}V^*_{0e}],
\end{array}
\end{equation}
which is generally non-vanishing. Here $u_{\alpha}$, $u_{\beta}$ and
$u_{\gamma}$ are the three up-type quarks. Therefore, the $d_e$ quark
EDM is proportional to,
\begin{equation}
2i\sum_{i,\alpha}{\rm Im}[V_{\alpha e}V^*_{\alpha i}V_{0i}V^*_{0e}]
H_i(\alpha).
\end{equation}
Note that to avoid complete cancellation due to
$\sum_i(V^*_{\alpha i}V_{0i})=\delta_{\alpha 0}=0$, $H_i(\alpha)$
must involve the $d_i$ mass.

Now let us evaluate analytically the $d_e$ quark EDM, $d(d_e)$.
This is facilitated by the mass hierarchy in the SM,
$m_t\gg m_W\gg m_q$, where $q$ stands for other five quarks, and the
assumption that $m_{d_4}\gg m_t$. We should discriminate two kinds of
down-type quarks with $d_4$ heavy and others light, as well as two
kinds of up-type quarks with top heavy and others light. We want to
retain only the terms that are least suppressed by light quark masses.
Since the charged current is purely left-handed, the chirality flip
needed for the EDM operator has to be made by the external quark mass.
We found that $H_i(\alpha)$ is proportional to $m^2_{u_{\alpha}}$ for
all of $WG$, $GW$ and $GG$ contributions when $u_{\alpha}$ is light.
Therefore we only need to keep the top quark in up-type quarks. The
leading terms involving the heavy $d_4$ quark come from the $WG$ and
$GG$ contributions:
\begin{equation}
\begin{array}{rcl}
\displaystyle d(d_e)_{\rm heavy}&=&em_{d_e}G^2_Fm^2_W(4\pi)^{-4}
{\rm Im}[V_{te}V^*_{t4}V_{04}V^*_{0e}]\\
\multicolumn{3}{r}{\left[Q_u\left(
\frac{\displaystyle 23}{\displaystyle 9}-
\frac{\displaystyle 8}{\displaystyle 9}\mu_t+
\frac{\displaystyle 4}{\displaystyle 3}\mu_t
\ln\frac{\displaystyle{\mu_4}}{\displaystyle{\mu_t}}-
\frac{\displaystyle 16}{\displaystyle 3}\ln\mu_t
\right)\right.}\\
\multicolumn{3}{r}{\left.+Q_d\left(
-\frac{\displaystyle 59}{\displaystyle 9}-
\frac{\displaystyle 1}{\displaystyle 3}\mu_t-
2\mu_t\ln\frac{\displaystyle{\mu_4}}{\displaystyle{\mu_t}}+
10\ln\mu_t
\right)\right]},
\end{array}
\end{equation}
where $\mu_t=m^2_t/m^2_W$ and $\mu_4=m^2_{d_4}/m^2_W$.
The leading terms involving the light $d_i$ quarks are independent of
their masses so that we may use
$\sum_iV^*_{\alpha i}V_{0i}=-V^*_{\alpha 4}V_{04}$
to sum up their contributions and obtain,
\begin{equation}
\begin{array}{rcl}
\displaystyle d(d_e)_{\rm light}&=&em_{d_e}G^2_Fm^2_W(4\pi)^{-4}
{\rm Im}[V_{te}V^*_{t4}V_{04}V^*_{0e}]\\
\multicolumn{3}{r}{\left[Q_u\left(
-8+\frac{\displaystyle{16\pi^2}}{\displaystyle 3}-
12\ln\mu_t+8\ln^2\mu_t
\right)\right.}\\
\multicolumn{3}{r}{\left.+Q_d\left(
-4-\frac{\displaystyle{ 8\pi^2}}{\displaystyle 3}+
4\ln\mu_t-8\ln^2\mu_t
\right)\right]},
\end{array}
\end{equation}
which comes from the $GW$ contribution.
We should emphasize that the above results are obtained in the limit
of large mass hierarchy so that the cancellation of the EDM in
the degeneracy limit of down-type quarks cannot be explicit
therein.

We note that the $d_e$ quark EDM, $d(d_e)$, occurs at order
$g^4(m_{d_e}/m_W^2)$ and there is no further suppression due to GIM
mechanism\cite{GIM}. This is typical for models without CKM unitarity.
Some terms are further enhanced by the heavy top mass, while the
absence of the heaviest $m_{d_4}^2$ enhancement is consistent with
general arguments based on gauge invariance and naive dimensional
analysis\cite{lkl}.

For numerical analysis, we use the following parameters:
$G_F=1.2\times 10^{-5}{\rm ~GeV}^{-2}$, $m_W=80$ GeV and $m_t=175$ GeV.
Then we have for the $d$ quark,
\begin{equation}
\begin{array}{rcl}
\displaystyle d(d)_{\rm heavy}&=&
{\rm Im}[V_{td}V^*_{t4}V_{04}V^*_{0d}]\cdot\frac{\displaystyle{ m_d}}
{\displaystyle{10~{\rm MeV}}}\\
\multicolumn{3}{r}{
\cdot (-5.3,-0.86,+2.3)\times 10^{-26}~{\rm e~cm}},\\
\multicolumn{3}{r}{{\rm for}~m_{d_4}=(200,300,400)~{\rm GeV}},\\
\displaystyle d(d)_{\rm light}&=&
{\rm Im}[V_{td}V^*_{t4}V_{04}V^*_{0d}]\cdot\frac{\displaystyle{m_d}}
{\displaystyle{10~{\rm MeV}}}\\
\multicolumn{3}{r}{\cdot 3.3\times 10^{-25}~{\rm e~cm}}.
\end{array}
\end{equation}
Note that the same combination of CKM elements is involved in the two
contributions. Since the `heavy' part is generally smaller by one order
of magnitude we retain below only the `light' part which is independent
of the $d_4$ mass. The product of CKM elements can be expressed in
terms of the $3\times3$ submatrix elements, e.g.,
\begin{equation}
{\rm Im}[V_{td}V^*_{t4}V_{04}V^*_{0d}]=
{\rm Im}[V_{td}V^*_{tb}Z_{bd}]-
{\rm Im}[V_{ts}V^*_{td}Z_{ds}],
\end{equation}
where
$Z_{ij}=V_{ui}V^*_{uj}+V_{ci}V^*_{cj}+V_{ti}V^*_{tj},~i,j=d,s,b$
are precisely the couplings appearing in the FCNC $Z$ interactions
amongst ordinary down-type quarks. The most stringent bounds on them
come from the neutral meson mixing and FCNC decays. Here we adopt the
bounds obtained by requiring that the new tree level FCNC
effects do not exceed the experimental values. Some bounds may be
relaxed if destructive interference occurs between them and other
contributions, e.g., the box diagrams. We take
$|Z_{ds}|\leq 3\times 10^{-4},~|Z_{sb}|\sim|Z_{bd}|\leq 10^{-3}$.
These bounds are actually interrelated with the extraction of other
elements like $V_{td}$ and $V_{ts}$. We do not attempt here a global
analysis which is beyond the main interest of the present work, but
simply adopt the following values for numerical estimate:
$|V_{td}|\sim 0.02, ~|V_{ts}|\sim|V_{cb}|\sim0.04.$
Then, ${\rm Im}[V_{td}V^*_{t4}V_{04}V^*_{0d}]\leq 2\times 10^{-5}$,
and
\begin{equation}
\begin{array}{l}
\displaystyle|d(d)_{\rm light}|\leq6.5\times 10^{-30}
\frac{m_d}{10~{\rm MeV}}~{\rm e~cm}.
\end{array}
\end{equation}
Using the $SU(6)$ relation for the neutron, we obtain
\begin{equation}
\displaystyle|d(n)|\leq 0.8\times 10^{-29}
\frac{m_d}{10~{\rm MeV}}~{\rm e~cm}.
\end{equation}

A few comments are in order.

(1) The non-unitarity of the CKM matrix generally induces FCNC
interactions of the $Z$ and Higgs bosons with down-type quarks. We found
that mixed exchanges of $W^{\pm}$ and $Z$ (or Higgs) can contribute
to the EDM. Since the Higgs boson decouples for heavy enough mass, we
consider here the FCNC $Z$ couplings. There are two types of diagrams.
The first one is similar to Fig.~\ref{C0b} and its leading term for the
$d(d)$ contains the same matrix elements as in Eq.(10) and decreases the
above result by about $1/3$ for $m_{d_4} = 300 {\rm GeV} $, while the
$d(u)$ is severely suppressed. The second type has a different topology
and involves the complicated issue of the renormalization of the
non-unitary CKM matrix. The details of all this will be reserved for a
future publication. The point we want to make here is that the purely
charged current result in Eq.(13) gives us the correct order of
magnitude of the neutron EDM since there is no plausible reason to
expect strong cancellation between the charged current and FCNC
contributions which generally involve several different Jarlskog rephasing
invariants of $CP$ violation.

(2) The above discussion can be easily generalized to other models
with exotic quarks. At the two loop level both up- and down-type
quark EDM's vanish strictly in the model with a sequential fourth
generation. In the model with an extra up-type singlet quark, the
down-type quark EDM's vanish identically at two loop order. Since
all down-type quarks are light in this case, the leading terms for
the $u_e$ EDM must be proportional to $m_{u_e}m^2_{d_i}$ and are thus
very small compared to the case considered above.

(3) We have also studied the $P$ and $T$ violating purely gluonic
operators, e.g., the dimension-$6$ Weinberg operator\cite{Wein}. We
found that they are severely suppressed by light quark masses in the
current case as in the SM \cite{ggg}. Their contribution to the
neutron EDM can be ignored. To obtain the quark chromoelectric dipole
moment(CEDM), one merely replaces $eQ_{u,d}$ by $g_s$ in Eqns.$(8)$
and $(9)$ with $g_s$ being the QCD coupling. It is then clear that
the quark CEDM is much smaller than the quark EDM so that the latter
remains to be the dominant contribution in the neutron EDM.

(4)In the static limit the fermionic part of the EDM is identical to
the spin operator. Since a significant amount of the proton spin is
derived from the polarized strange quark sea\cite{ll96}, it seems
reasonable that the strange quark also contributes to the neutron
EDM\cite{ZK}. For the strange quark EDM, we have enhancement factors
from masses and CKM elements. For the latter the dominant one is
${\rm Im}[V_{tb}V_{ts}^*Z_{sb}]$. Suppose $|\eta|\sim 10\%$ of the
proton spin is accounted for by the strange sea, then we have
$|d(n)_{\rm strange}|\leq 0.8\times 10^{-29}(m_s/m_d)
(|V_{ts}|/|V_{td}|)|\eta|~{\rm e~cm}\sim 3.2\times 10^{-29}~{\rm e~cm}$
for the quoted parameters and $m_s=200$ MeV.

(5) Another possible enhancement for the neutron EDM originates from
long distance physics\cite{distance}. Although there are still controversies
concerning this, it seems reliable to get an enhancement factor of two
orders of magnitude in the SM. If this persists in the case considered
here, improvement of the upper bound on the neutron EDM in the near
future will provide an interesting test of the unitarity of the CKM
matrix which will be complementary to or even competitive with the
bound from $B$ physics.

The work of X. Li was supported in part by the China National Natural
Science Foundation under grant Numbers 19835060 and 19875072.

{\tighten

} 
\end{multicols}
\end{document}